\documentclass[12pt]{article}
\usepackage{amsmath}
\usepackage{graphicx,psfrag,epsf}
\usepackage{enumerate}
\usepackage{natbib}
\usepackage{url} 
\usepackage{amsfonts,amssymb,latexsym, amsmath}
\usepackage{amsthm}  
\usepackage{graphicx}
\usepackage{natbib}
\usepackage{wrapfig}
\usepackage{lstautogobble}
\usepackage{multicol}
\usepackage{natbib}
\usepackage{threeparttablex}
\usepackage{booktabs}
\usepackage{pdflscape} 
\usepackage{hyperref} 
\setcitestyle{authoryear,open={(},close={)}} 
\usepackage{sistyle} 
\allowdisplaybreaks
\newcommand{\rev}[1]{{\color{black}#1}}%

\usepackage[svgnames]{xcolor}
\usepackage{listings}
\newcommand{\blind}{0}
\newcommand{\bbeta}{\pmb{\beta}} 
\newcommand{\bbetahat}{\hat{\pmb{\beta}}} 
\newcommand{\bz}{\pmb{Z}} 

\usepackage{titlesec}

\titlespacing\section{0pt}{12pt plus 4pt minus 2pt}{0pt plus 2pt minus 2pt}
\titlespacing\subsection{0pt}{12pt plus 4pt minus 2pt}{0pt plus 2pt minus 2pt}
\titlespacing\subsubsection{0pt}{12pt plus 4pt minus 2pt}{0pt plus 2pt minus 2pt}

\begin{document}

\def\spacingset#1{\renewcommand{\baselinestretch}%
{#1}\small\normalsize} \spacingset{1}

\if0\blind
{
  \title{\bf Linking Potentially Misclassified Healthy Food Access to Diabetes Prevalence} 
  \author{Ashley E. Mullan\thanks{
    The authors gratefully acknowledge the Andrew Sabin Family Center for Environment and Sustainability for a seed grant that supported this work; the DEAC computing cluster team at Wake Forest University; and Lucy D'Agostino McGowan, David Kline, and Staci Hepler for their support when this work was once a thesis project.}\hspace{.2cm},
    P. D. Anh Nguyen, 
    and Sarah C. Lotspeich \\
    Department of Statistical Sciences, Wake Forest University}
  \maketitle
} \fi

\if1\blind
{
  \bigskip
  \bigskip
  \bigskip
  \begin{center}
    {\LARGE\bf Linking Potentially Misclassified Healthy Food Access to Diabetes Prevalence}
\end{center}
  \medskip
} \fi

\bigskip
\begin{abstract}
Access to healthy food is key to maintaining a healthy lifestyle and can be quantified by the distance to the nearest grocery store. However, calculating this distance forces a trade-off between cost and correctness. Accurate route-based distances following passable roads are cost-prohibitive, while simple straight-line distances ignoring infrastructure and natural barriers are accessible yet error-prone. Categorizing low-access neighborhoods based on these straight-line distances induces misclassification and introduces bias into standard regression models  estimating the relationship between disease prevalence and access. Yet, fully observing the more accurate, route-based food access measure is often impossible, which induces a missing data problem. We combat bias and address missingness with a new maximum likelihood estimator for Poisson regression with a binary, misclassified exposure (access to healthy food within some threshold), where the misclassification may depend on additional error-free covariates. In simulations, we show the consequence of ignoring the misclassification (bias) and how the proposed estimator corrects for bias while preserving more statistical efficiency than the complete case analysis (i.e., deleting observations with missing data). Finally, we apply our estimator to model the relationship between census tract diabetes prevalence and access to healthy food in northwestern North Carolina.
\end{abstract}

\noindent%
{\it Keywords:} Food Environment, Maximum Likelihood Estimation, Two-Phase Design, Missing Data, One-Sided Misclassification, Poisson Regression 
\vfill

\newpage
\spacingset{1.45} 
\section{Introduction}
\label{sec:intro}

Maintaining a healthy diet is more than simply a choice. A person must have consistent access to and ability to obtain healthy food. 
Physical access to healthy foods may be hindered by many factors, including the neighborhood's geography or community's socioeconomic makeup \citep{wolfson2019}. 
Subsequently, neighborhoods with poor food access may see higher prevalences of diet-related diseases  \citep{gucciardi2014}. \rev{For example, \cite{lopez2007} links food access, specifically the presence of at least one grocery store within a zip code, to a 10\% decrease in obesity risk when compared to zip codes without this access. Groups such as food policy councils \citep{lange2021associations} prioritize health equity by supporting policies that increase access to healthy food in rural and minority communities.} We focus on the relationship between diabetes prevalence and food access at the neighborhood (census tract) level, although our methods adapt to other outcomes and units as well. 

To estimate this relationship, we must first quantify food access, for which there are multiple metrics. 
We devote our attention to ``access indicators,'' which classify neighborhoods into binary categories of having/not having access to healthy food within some fixed distance. The United States Department of Agriculture (USDA) includes indicators like these in their Food Research Atlas \citep{atlas}.
Computing these food access indicators requires researchers to first calculate the pairwise distances between neighborhoods and potential healthy food retailers. Just like there are multiple ways to measure food access, there are multiple ways to measure these distances. For instance, the Haversine formula  modifies the traditional straight-line Euclidean distance between two points (e.g., a neighborhood centroid and grocery store) to account for the curvature of the Earth. Alternatively, route-based distance calculations are possible, which only allow travel via established infrastructure (like roads) to connect the two points. 

These methods each have their advantages, and choosing between them to calculate distances requires a trade-off. The Haversine formula is a simple mathematical equation, making it free, easy, and quick to calculate in existing software \citep[e.g.,][]{geosphere}. However, it underestimates distances to healthy food retailers and therefore overestimates neighborhoods' food access. This underestimation stems from the fact that straight-line distances are necessarily less than or equal to any other measure of distance. In our study, underestimated Haversine distances can lead to neighborhoods being misclassified as having access to healthy foods when they do not (false positives). Specifically, the access indicators are subject to one-sided misclassification; false positives are the only possible type of error. 

In short, Haversine distance is easily obtained but leads to access indicators that may be incorrect. Taking route-based distances via software like Google Maps circumvents the possibility of unrealistic, impassable routes. Still, these more-accurate distances are computationally expensive and, despite availability of open source options, may also be costly. Therefore, calculating them for all neighborhoods may not be plausible. 

Luckily, some accurate data are better than none,  so we frame our study as a two-phase design. We measure the diabetes prevalence and the error-prone food access indicator based on Haversine distances for all neighborhoods (Phase I). Then, we query a subset of neighborhoods using Google Maps \citep{ggmap} to obtain the route-based distances for the more-accurate food access indicator (Phase II). We leverage missing data techniques to include every neighborhood in the model in some way, and this strategy yields unbiased estimates of the exposure effects (food access) on the outcome (diabetes prevalence) without fully observing the route-based exposures. For extensive reviews of statistical methods for error-prone data, see \citet{Keogh2020} and \citet{Shaw2020}. 

For our study, with diabetes prevalence as the outcome and (potentially mismeasured) food access indicator as the exposure, a new maximum likelihood estimator (MLE) for Poisson regression with exposure misclassification is needed. \rev{This MLE incorporates all information for all neighborhoods, offering better statistical efficiency than design-based approaches, like inverse probability weighting. As a model-based approach, it offers flexibility in choosing the queried subset; design-based approaches require nonzero sampling probability for all observations. Moreover, it is computationally faster than  multi-stage model-based approaches, like multiple imputation or Bayesian modeling, and it would be straightforward to adapt for more robustness (e.g., semiparametric MLE).} The rest of this paper discusses the derivation of the MLE (Section~\ref{chap:methods}), extensive simulations  (Section~\ref{chap:sims}), and a case study (Section~\ref{ch:case_study}). We conclude by discussing findings and next steps (Section~\ref{ch:concl}). 

\section{Methods}\label{chap:methods}

\subsection{Model and Notation}

We model $Y$, a count outcome offset by $O$, given a binary exposure  $X$ and covariate vector $\bz$. An error-prone, potentially misclassified version of the exposure, $X^*$, is available for all $N$ sampled neighborhoods, while true $X$ is only available for a subset of $n$ ``queried'' ones ($n < N$). 
We write the joint probability of a complete (queried) observation as follows:
\begin{align}
{\Pr}_{\bbeta,\pmb{\eta}}(Y, X, X^*, \bz) &= {\Pr}(Y \mid X, X^*, \bz){\Pr}_{\pmb{\eta}}(X \mid X^*, \bz){\Pr}(X^*, \bz) \nonumber \\
&= {\Pr}_{\bbeta}(Y \mid X, \bz){\Pr}_{\pmb{\eta}}(X \mid X^*, \bz){\Pr}(X^*, \bz). 
\label{p4}
\end{align}

First, ${\Pr}_{\bbeta}(Y \mid X, \bz)$ denotes the Poisson regression for $Y$ given ($X$, $\bz$), parameterized by log prevalence ratios (log PRs) $\bbeta$. Second, ${\Pr}_{\pmb{\eta}}(X \mid X^*, \bz)$ denotes the conditional probability mass function for $X$ given ($X^*$, $\bz$). We estimate this \textit{misclassification mechanism} using logistic regression, governed by $\pmb{\eta}$. To accommodate one-sided misclassification (e.g., only false positives), this model must be modified (Web Appendix A.1) but the rest of the methods remain unchanged. Third,  ${\Pr}(X^*, \bz)$ denotes the joint probability density or mass function of  ($X^*$,$\bz$). Since ($X^*$,$\bz$) are fully observed, ${\Pr}(X^*, \bz)$ drops out of our log-likelihood for $\bbeta$, so we can leave this model unspecified. Notice that the second equality in \eqref{p4} assumes conditional independence of  ($Y$, $X^*$) given ($X$, $\bz$) (i.e., surrogacy). 

In our setting, all $N$ neighborhoods have an observed $X^*$ based on Haversine distances, but only a limited subset of size $n$ has true $X$ measured based on route-based distances. Thus, the data available for a neighborhood depend on whether it was queried:
\begin{align*}
    \text{Observed data} &= 
    \begin{cases} 
    (Y, X, X^*, \bz) & \text{ if the neighborhood was queried, and } \\
   (Y, X^*, \bz) & \text{ otherwise.}
   \end{cases}
\end{align*}
Queried neighborhoods are assumed to follow 
\eqref{p4}, while unqueried ones are assumed to instead follow: 
\begin{align*}
{\Pr}_{\bbeta, \pmb{\eta}}(Y,X^*,\bz) &= \sum_{x=0}^{1}{\Pr}_{\bbeta}(Y \mid X=x, \bz){\Pr}_{\pmb{\eta}}(X=x \mid X^*, \bz){\Pr}(X^*, \bz). 
\end{align*}

\subsection{Maximum Likelihood Estimation}

We piece the queried and unqueried observations and their joint probability functions together, along with an indicator $Q_i$ of queried status, to define the likelihood function: 
\begin{align}
   \mathcal{L}_N(\bbeta, \pmb{\eta}) &= \prod_{i=1}^{N} \{{\Pr}_{\bbeta, \pmb{\eta}}(X_i,X_i^*,Y_i,\bz_i)\}^{Q_i}\{{\Pr}_{\bbeta, \pmb{\eta}}(X_i^*,Y_i,\bz_i)\}^{1 - Q_i}\nonumber\\
   &\propto \prod_{i=1}^{N} \{{\Pr}_{\bbeta}(Y_i \mid X_i, \bz_i){\Pr}_{\pmb{\eta}}(X_i \mid X_i^*, \bz_i)\}^{Q_i}\nonumber \\
   &\phantom{\propto \prod_{i=1}^{N}} \times \left\{\sum_{x=0}^{1}{\Pr}_{\bbeta}(Y_i \mid X_i=x, \bz_i){\Pr}_{\pmb{\eta}}(X_i=x \mid X_i^*, \bz_i)\right\}^{1 - Q_i} . \label{lik:c4}
\end{align}
The query indicator controls which probability function governs observation $i$ in \eqref{lik:c4}; it evaluates to $1$ for queried neighborhoods (i.e., when $X_i$ is available) and to $0$ otherwise. Notice that the $N$ neighborhoods are assumed to be independent in this likelihood. Modifying the likelihood to allow spatial autocorrelation is a promising future direction.

The MLEs $\bbetahat$ maximize \eqref{lik:c4} and are found via an EM algorithm (Web Appendix A.2 and Supplemental Figure~S1). Standard errors (SEs) for $\bbetahat$ are conveniently estimated using numerical derivatives. Our proposed MLEs are implemented in the \textit{possum} R package, available at \url{www.github.com/sarahlotspeich/possum}. Under regularity conditions, missingness at random, and proper specification of both models, the MLEs are asymptotically consistent, normal, and efficient.

\section{Simulations}\label{chap:sims}

We conducted simulation studies to test the performance of our MLEs against the gold standard, naive, and complete case analyses. The gold standard analysis fit a Poisson regression using $X$ for all observations, while the naive analysis used error-prone version $X^*$, and the complete case used $X$ for only the queried subset. Data generation was designed to mimic our data from the Piedmont Triad (Section~\ref{ch:case_study}), with primary variations of the sample size $N$, query proportion $q$, and positive predictive value $PPV = \Pr(X=1|X^*=1,Z)$. 

We considered samples of $N = 390$ neighborhoods, approximately the number of census tracts in the Piedmont Triad, or $N = 2200$ neighborhoods, approximately the number of census tracts in North Carolina. In all settings, we first simulated a covariate ($Z$), followed by a potentially misclassified exposure ($X^* \mid Z$), a true exposure ($X \mid X^*,Z$), an offset ($O$), and a count outcome ($Y \mid X, Z$). Nuisance parameters were chosen to force a prespecified $PPV$. Based on query proportion $q$ ($q \in [0,1]$), a subset of $n = Nq$ neighborhoods had non-missing $X$ for the complete case and MLE analyses. All $n$ were chosen from those with $X^* = 1$. Since false negatives for food access are not possible, neighborhoods with $X^* = 0$ must have $X = 0$ and thus do not need to be queried. For more details on simulation setup, see Web Appendix B.1. We ran \num{1000} replications per setting and recorded summary metrics capturing empirical bias, variance, and relative efficiency for $\hat{\beta}_1$, the adjusted log PR on $X$. We also assessed the empirical coverage probability (CP) of the 95\% Wald confidence interval and the validity of the SE estimator.

We focus here on results under varied misclassification rates (Section~\ref{subsec:sim_misclass}) and query percentages (Section~\ref{subsec:sim_q}). Unless otherwise stated, we fixed $q = 0.1$, $PPV = 0.6$, $\beta_1 = 0.18$, and outcome prevalence $\approx 10\%$. Additional results with two-sided misclassification in $X^*$ and varied outcome prevalence can be found in Web Appendices~B.2--B.3. Like the settings below, the MLE remained unbiased and more efficient than the complete case analysis. 

\subsection{Results Under Varied Misclassification Rates}\label{subsec:sim_misclass}
We considered $PPV$ from $0.5$ to $0.9$ at sample sizes of $N = 390$ and $2200$. 
The bias of the MLE remained comparable to that of the gold standard and complete case analyses in all settings (Table~\ref{sims:ppv}). Even under the most severe misclassification ($PPV = 0.5$), the MLE was less than  $0.1\%$ biased. However, the naive analysis yielded heavy downward bias ranging from $-9\%$ to $-47\%$, worsening as $PPV$ decreased  and persisting in larger samples. 

\begin{table}[h]
\centering
\caption{Simulation results for $\hat{\beta}_1$, the adjusted log PR on $X$, under varied positive predictive value ($PPV$) and sample size ($N$), assuming $\approx 10\%$ outcome prevalence, $\beta_1 = 0.18$, and $q =0.1$ query percentage.}
\resizebox{\columnwidth}{!}{
\begin{threeparttable}
\centering
\begin{tabular}{ccrcrcrccrcccc}
\toprule
\multicolumn{2}{c}{\textbf{ }} & \multicolumn{2}{c}{\textbf{Gold Standard}} & \multicolumn{2}{c}{\textbf{Naive}} & \multicolumn{3}{c}{\textbf{Complete Case}} & \multicolumn{5}{c}{\textbf{Maximum Likelihood Estimator}} \\
\cmidrule(l{3pt}r{3pt}){3-4} \cmidrule(l{3pt}r{3pt}){5-6} \cmidrule(l{3pt}r{3pt}){7-9} \cmidrule(l{3pt}r{3pt}){10-14}
\textbf{$\pmb{N}$} & \textbf{$\pmb{PPV}$} & \textbf{Bias} & \textbf{ESE} & \textbf{Bias} & \textbf{ESE} & \textbf{Bias} & \textbf{ESE} & \textbf{RE} & \textbf{Bias} & \textbf{ESE} & \textbf{ASE} & \textbf{CP} & \textbf{RE}\\
\midrule
390 & 0.5 & $0.000$ & $0.005$ & $-0.465$ & $0.008$ & $ 0.002$ & $0.015$ & $0.098$ & $0.000$ & $0.005$ & $0.005$ & $0.949$ & $0.955$\\
 & 0.6 & $0.000$ & $0.005$ & $-0.367$ & $0.007$ & $ 0.000$ & $0.016$ & $0.085$ & $0.000$ & $0.005$ & $0.005$ & $0.954$ & $0.975$\\
 & 0.7 & $0.001$ & $0.005$ & $-0.271$ & $0.007$ & $-0.001$ & $0.018$ & $0.075$ & $0.001$ & $0.005$ & $0.005$ & $0.938$ & $0.985$\\
 & 0.8 & $0.000$ & $0.005$ & $-0.181$ & $0.007$ & $ 0.000$ & $0.019$ & $0.059$ & $0.000$ & $0.005$ & $0.005$ & $0.953$ & $0.985$\\
 & 0.9 & $0.000$ & $0.005$ & $-0.089$ & $0.006$ & $ 0.000$ & $0.030$ & $0.028$ & $0.000$ & $0.005$ & $0.005$ & $0.944$ & $0.987$\\
\addlinespace
2200 & 0.5 & $ 0.000$ & $0.002$ & $-0.465$ & $0.003$ & $-0.001$ & $0.006$ & $0.107$ & $ 0.000$ & $0.002$ & $0.002$ & $0.943$ & $0.986$\\
 & 0.6 & $ 0.001$ & $0.002$ & $-0.366$ & $0.003$ & $ 0.001$ & $0.007$ & $0.097$ & $ 0.001$ & $0.002$ & $0.002$ & $0.935$ & $0.984$\\
 & 0.7 & $ 0.000$ & $0.002$ & $-0.272$ & $0.003$ & $-0.002$ & $0.007$ & $0.087$ & $ 0.000$ & $0.002$ & $0.002$ & $0.951$ & $0.992$\\
 & 0.8 & $ 0.000$ & $0.002$ & $-0.179$ & $0.003$ & $ 0.000$ & $0.008$ & $0.065$ & $ 0.000$ & $0.002$ & $0.002$ & $0.948$ & $1.002$\\
 & 0.9 & $-0.001$ & $0.002$ & $-0.089$ & $0.003$ & $-0.002$ & $0.011$ & $0.036$ & $-0.001$ & $0.002$ & $0.002$ & $0.944$ & $0.992$\\
\bottomrule
\end{tabular}
\begin{tablenotes}[flushleft]
\item{\em Note:} Bias and ESE are, respectively, the empirical relative bias and standard error of each estimator; ASE is the average of the standard error estimator; CP is the empirical coverage probability of the \textcolor{black}{95\% Wald confidence intervals (computed with the standard error estimator)}; and RE is the relative efficiency of the estimator to the gold standard. All entries are based on \num{1000} replicates. Across all settings, 22 replicates of the MLE ($0.2\%$) were replaced with the naive analysis due to $100\%$ PPV in the queried subset and perfect separation in the misclassification model.
\end{tablenotes}
\label{sims:ppv}
\end{threeparttable}}
\end{table}

The empirical SEs of the MLE were smaller than those of the complete case analysis in every setting. In addition, the average SEs of the MLE approximated the empirical versions very well, which also led to empirical CP in the correct neighborhood of 95\%. Finally, the relative efficiency of the MLE to the gold standard analysis was over five times higher than that of the complete case analysis in every setting (between $0.96$--$1.00$ versus $0.04$--$0.11$). 

\subsection{Results Under Varied Query Percentages}\label{subsec:sim_q}

We also considered querying proportions between $q = 0.1$ and $0.5$ from samples of $N = 390$ and $2200$ neighborhoods. The bias of the MLE remained comparable to that of the gold standard analysis; both methods were under 0.5\% biased in all settings (Table~\ref{sims:q}). The empirical SEs of the MLE continued to match or beat those from the complete case analysis in every setting. With the smallest $q$, the SEs from the complete case analysis were over three times the size of the MLE's. Again, the average SEs of the MLE matched their empirical counterparts, and empirical CP was close to the nominal 95\% level. 

\begin{table}[ht]
\centering
\caption{Simulation results for $\hat{\beta}_1$ under varied query percentage ($q$) and sample size ($N$), assuming $\approx 10\%$ outcome prevalence, $\beta_1 = 0.18$, and positive predictive value $= 0.6$.}
\resizebox{\columnwidth}{!}{
\begin{threeparttable}
\begin{tabular}{ccrcrcrccrcccc}
\toprule
\multicolumn{2}{c}{\textbf{ }} & \multicolumn{2}{c}{\textbf{Gold Standard}} & \multicolumn{2}{c}{\textbf{Naive}} & \multicolumn{3}{c}{\textbf{Complete Case}} & \multicolumn{5}{c}{\textbf{Maximum Likelihood Estimator}} \\
\cmidrule(l{3pt}r{3pt}){3-4} \cmidrule(l{3pt}r{3pt}){5-6} \cmidrule(l{3pt}r{3pt}){7-9} \cmidrule(l{3pt}r{3pt}){10-14}
\textbf{$\pmb{N}$} & \textbf{$\pmb{q}$} & \textbf{Bias} & \textbf{ESE} & \textbf{Bias} & \textbf{ESE} & \textbf{Bias} & \textbf{ESE} & \textbf{RE} & \textbf{Bias} & \textbf{ESE} & \textbf{ASE} & \textbf{CP} & \textbf{RE}\\
\midrule
390 & 0.10 & $0.000$ & $0.005$ & $-0.367$ & $0.007$ & $0.000$ & $0.016$ & $0.085$ & $0.000$ & $0.005$ & $0.005$ & $0.954$ & $0.975$\\
 & 0.25 & $0.002$ & $0.005$ & $-0.367$ & $0.007$ & $0.003$ & $0.010$ & $0.227$ & $0.002$ & $0.005$ & $0.005$ & $0.953$ & $0.974$\\
 & 0.50 & $0.001$ & $0.005$ & $-0.366$ & $0.007$ & $0.002$ & $0.007$ & $0.463$ & $0.001$ & $0.005$ & $0.005$ & $0.953$ & $0.992$\\
\addlinespace
2200 & 0.10 & $ 0.001$ & $0.002$ & $-0.366$ & $0.003$ & $ 0.001$ & $0.007$ & $0.097$ & $ 0.001$ & $0.002$ & $0.002$ & $0.935$ & $0.984$\\
 & 0.25 & $ 0.000$ & $0.002$ & $-0.368$ & $0.003$ & $ 0.000$ & $0.004$ & $0.227$ & $ 0.000$ & $0.002$ & $0.002$ & $0.949$ & $0.980$\\
 & 0.50 & $ 0.000$ & $0.002$ & $-0.367$ & $0.003$ & $ 0.000$ & $0.003$ & $0.508$ & $ 0.000$ & $0.002$ & $0.002$ & $0.945$ & $0.992$\\
\bottomrule
\end{tabular}
\begin{tablenotes}[flushleft]
\item{\em Note:} Bias and ESE are, respectively, the empirical relative bias and standard error of each estimator; ASE is the average of the standard error estimator; CP is the empirical coverage probability of the \textcolor{black}{95\% Wald confidence intervals (computed with the standard error estimator)}; and RE is the relative efficiency of the estimator as compared to the gold standard. All entries are based on \num{1000} replicates. 
\end{tablenotes}
\end{threeparttable}
}
\label{sims:q}
\end{table}

The relative efficiency of the MLE to the gold standard analysis exceeded that of the complete case analysis in every setting. This trend was especially noticeable (as expected) when $q$ was smaller, when larger percentages of observations were missing error-free exposures. The complete case analysis would have thrown these observations away.

\section{Case Study: Piedmont Triad, North Carolina} \label{ch:case_study}

\subsection{Data and Objectives}

We investigated the associations between having access to healthy foods and diabetes prevalence across $N = 387$ census tracts in the Piedmont Triad region of northwestern North Carolina. A census tract is a small, relatively permanent statistical subdivision of a county or statistically equivalent entity \citep{tracts}. Approximately 1.7 million people live in the Piedmont Triad, and the region is known to have excellent transportation infrastructure. 
The model of interest is defined as $\log\left\{\textrm{E}_{\bbeta}(Y\mid X,Z)\right\} = \beta_0 + \beta_1 X + \beta_2 Z+ \beta_3 X \times Z + \log(O)$, where $Y$ is the number of diabetes cases, $O$ is the population (used as an offset), $X$ is an indicator of having access to healthy foods within a given radius, and $Z$ is an indicator of metropolitan status. We calculated the Haversine and route-based distances from the population-weighted centroid, according to the 2020 U.S. Census, to the nearest healthy food retailer. Using one-mile and half-mile radii, we discretized these distances to create our exposures $X$ and $X^*$. For example, $X = 1$ if a census tract had at least one healthy food retailer within the threshold and $=0$ otherwise.

Queried data (i.e., with both $X$ and $X^*$) were available for all census tracts. However, to demonstrate the two-phase design, we assumed that only a subset had available route-based access data for the MLEs and complete case analyses. After running the gold standard analysis on the full data, we created a partially queried version wherein $X$ was only available for census tracts with $X^* = 0$ (such that $X = 0$ necessarily) and a subset of those with $X^* = 1$. A random sample of $n = 77$ tracts with $X^* = 1$, targeting a 50/50 split between metropolitan/non-metropolitan, was chosen \rev{(Supplemental Figure~S2)}. Only tracts with error-prone access (i.e., $X^* = 1$) needed to be validated, since those with $X^*=0$ must have $X=0$ and do not inform the misclassification mechanism. The naive analysis used the error-prone access variable on the entire sample, and the complete case analysis used error-free access variable on the subset of queried data. The MLEs used the error-prone and error-free access indicators when available and just the error-prone version otherwise.

\subsection{Diabetes Prevalence and Healthy Food Access}

\begin{figure}[ht]
\begin{center}
\includegraphics[width=0.7\textwidth]{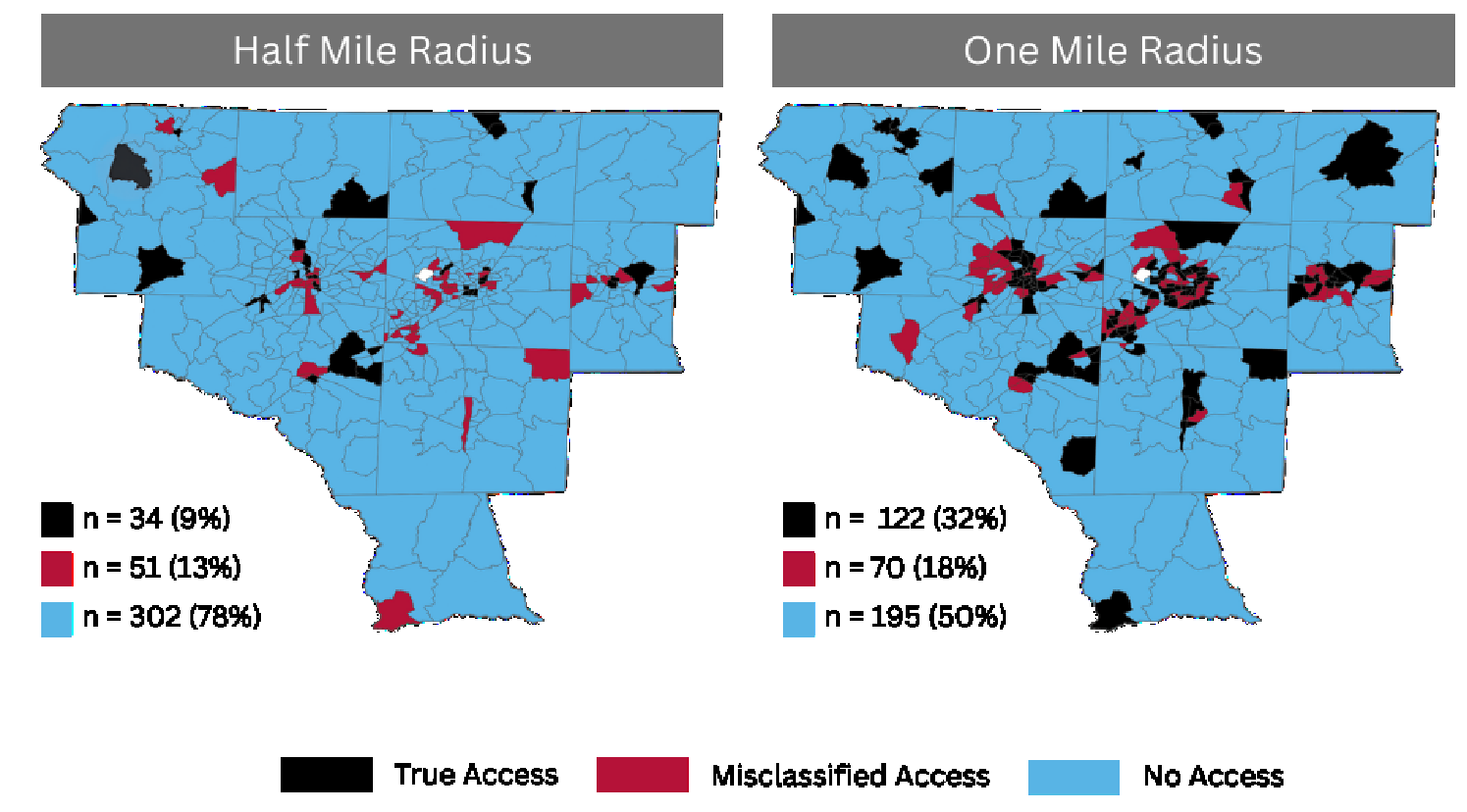}
\end{center}
\caption{Map of Piedmont Triad census tracts' access  at one-half and one mile thresholds.
\label{fig:landscape_maps}}
\vspace*{-8mm}
\end{figure}

At both half- and one-mile radii, most tracts ($n = 353$ [$91\%$] and $n = 265$ [$68\%$], respectively) did not have access to healthy foods (Figure~\ref{fig:landscape_maps}). Larger, rural tracts tended not to have access. Smaller, urban tracts tended to have access, or at least to be misclassified as such based on Haversine distances. Of the $192$ tracts with access to healthy foods based on one-mile Haversine distances, only $122$ actually had it based on route-based ones ($PPV = 64\%$). As expected, the smaller threshold saw more misclassification ($PPV = 40\%$). 

The estimated  $\widehat{\textrm{PR}}_{non} = \exp(\hat{\beta_1})$ compared the expected diabetes prevalence in non-metropolitan census tracts with versus without access to healthy foods. 
In metropolitan census tracts, $\widehat{\textrm{PR}}_{metro} = \exp(\hat{\beta_1} + \hat{\beta_3})$ was interpreted similarly, and its 95\% CI calculated as $\exp\left\{(\hat{\beta_1} + \hat{\beta_3}) \pm 1.96 \widehat{\textrm{SE}}(\hat{\beta_1} + \hat{\beta_3})\right\}$, where $\widehat{\textrm{SE}}(\hat{\beta_1} + \hat{\beta_3}) = \sqrt{\widehat{\textrm{V}}(\hat{\beta_1}) + \widehat{\textrm{V}}(\hat{\beta_3}) + \widehat{\textrm{Cov}}(\hat{\beta_1}, \hat{\beta_3})}$. We focus on these two PRs; the full models are summarized in Supplemental Table~S4. \rev{Note that these PRs should be interpreted through an inferential framework; they could also be used for prediction, but we cannot conclude that the relationships are causal.}  

At either threshold, diabetes prevalence in neighborhoods with access to healthy foods was expected to be higher than in those without (Figure~\ref{fig:forest_b1}). \rev{This trend seems somewhat counterintuitive, but the literature does not agree on a clear pattern linking food access and diabetes (e.g., Gucciardi et al., 2014 versus Flint et al., 2020). Potential confounding (e.g., by socioeconomic status) or uncertainty around the outcome data (small area estimates) may play a role in this debate. 

\begin{figure}[ht]
\caption{Prevalence ratio estimates and 95\% confidence intervals for healthy food access at a one-half and one mile threshold in the Piedmont Triad using four analysis methods. 
\label{fig:forest_b1}}
\begin{center}
\vspace*{-2mm}
\includegraphics[width=5in]{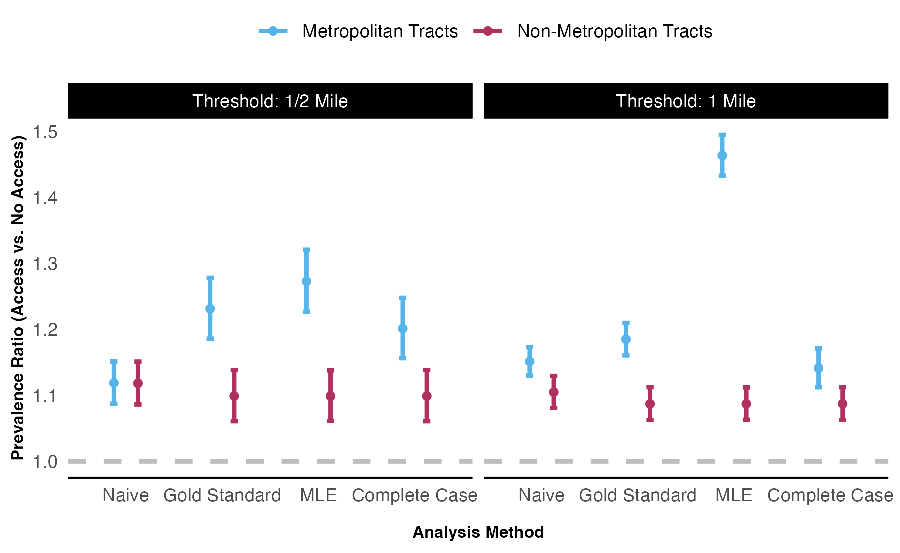}
\vspace*{-8mm}
\end{center}
\end{figure}

The observed effect direction} 
held at both distance thresholds and across all analysis methods. For example, based on the gold standard, diabetes prevalence in metropolitan tracts with access to healthy foods at one mile was expected to be 18\% higher than in those without ($\widehat{\textrm{PR}}_{metro} =1.18$; 95\% CI: $1.16$, $1.21$); among non-metropolitan tracts, those with access were expected to have 9\% higher prevalence ($\widehat{\textrm{PR}}_{non} =1.09$; 95\% CI: $1.06$, $1.11$). 

The difference between tracts with and without access was more pronounced in metropolitan areas, according to the gold standard analyses. At both thresholds, the MLE also captured this difference, while the complete case analysis only did so at the one-half mile radius, and the naive analysis  did not catch it at either threshold. The MLEs actually exaggerated this difference at the one-mile threshold but otherwise closely aligned with the gold standard analysis. Its 95\% confidence intervals were slightly narrower than the complete case. For more details and further sensitivity analysis, see Web Appendix C.

\section{Conclusion}\label{ch:concl}
We introduced a novel maximum likelihood estimator for Poisson regression with a misclassified binary exposure. In simulations, the MLE, on average, corrected for bias induced by the misclassified exposure in the naive analysis and reduced variance induced by deleting unqueried data in the complete case analysis. 
In our analysis of the Piedmont Triad, census tracts with access to healthy foods had higher diabetes prevalence, on average, than those without. This difference was more pronounced for metropolitan tracts. 

Our approach maximizes available information and addresses some of the concerns that arise in competing approaches. 
In simulations, the MLE had lower bias than the naive analysis, and it recovered much of the efficiency lost by the complete case analysis. For broader applicability, our MLE handles both one- and two-sided exposure misclassification. 

We are potentially limited by our choice to assume a parametric logistic regression model for the exposure error mechanism. However, this model has been shown to be fairly robust to misspecification  \citep{lotspeich2022efficient}. Still, nonparametric or semiparametric error models are promising. Results from the case study are limited by the nature of the available data. Census tracts do not always fairly describe neighborhoods, but they were the most granular level of available diabetes data. \rev{Other studies \citep[e.g.,][]{rummo2021} consider food access at the individual level. If we had individual-level data, our proposed methods would still apply, simply omitting the model offset.} Additionally, the model used to generate the small area diabetes estimates incorporates many socioeconomic factors that may have been relevant to our analysis but could not be included due to ``double dipping.'' 

In the future, it may be informative to explore spatial relationships between adjacent tracts, vary the outcome model, or modify the MLE to handle other data types. The MLE's efficiency could be further improved by a more strategic query design. Targeted strategies to select the most informative observations are common in the two-phase design literature. 

\phantom{\citep{flint2020}} 

\bigskip
\begin{center}
{\large\bf SUPPLEMENTARY MATERIAL}
\end{center}

\begin{description}

\item[Supporting information:] Web Appendices and Supplemental Figures and Tables (.pdf)

\item[Data:] Available at \url{https://github.com/sarahlotspeich/food_access_imputation}.

\item[R code:] R package \textit{possum} implementing the MLE (GNU zipped tar file) and R scripts for the simulation studies and Piedmont Triad analysis (.zip file)

\item[Additional materials: ] Additional code and results available for reproducibility at \\ \url{https://github.com/ashleymullan/food_access_misclassification}.

\end{description}

\bibliographystyle{Chicago}

\bibliography{ref}
\end{document}


\def\spacingset#1{\renewcommand{\baselinestretch}%
{#1}\small\normalsize} \spacingset{1}


\if0\blind
{
  \title{\bf Supplementary Materials for ``Linking Potentially Misclassified Healthy Food Access to Diabetes Prevalence''} 
  \author{Ashley E. Mullan,
    P. D. Anh Nguyen, 
    and 
    Sarah C. Lotspeich}
  \maketitle
} \fi

\if1\blind
{
  \bigskip
  \bigskip
  \bigskip
  \begin{center}
    {\LARGE\bf Supplementary Materials for ``Linking Potentially Misclassified Healthy Food Access to Diabetes Prevalence''} 
\end{center}
  \medskip
} \fi



\spacingset{1.45} 
\section{Additional Details About the Maximum Likelihood Estimator}

\subsection{One-Sided Misclassification}

In our case study, the error-prone access indicator $X^*$ was derived from straight-line distances and the error-free version $X$ is derived from route-based ones. By definition, the straight-line distances between all neighborhoods and healthy food retailers must be less than or equal to the corresponding route-based ones. Therefore, any neighborhood that does not have error-prone access (i.e., $X^* = 0$) must not have route-based access (i.e., $X = 0$). As a result, misclassification in $X^*$ is one-sided, because only false positives (i.e., observations with $X^* = 1$ but $X = 0$) are possible.  

Since $X$ is binary, we estimate the \textit{misclassification mechanism} ${\Pr}_{\pmb{\eta}}(X \mid X^*, \bz)$ using logistic regression, parameterized by $\pmb{\eta}$. As written (here and in Section~2.1), this model is appropriate for two-sided misclassification, wherein both false positives and false negatives are possible. For one-sided misclassification, as in our Piedmont Triad analysis, this model needs to be modified as follows:
\begin{align}
    {\Pr}_{\pmb{\eta}}(X \mid X^*, \bz) &= 
    \begin{cases} 
    {\Pr}_{\pmb{\eta}}(X \mid \bz) & \text{ if } X^* = 1 \\
    \textrm{I}(X = 0) & \text{ if } X^* = 0,
    \end{cases}
\end{align}
where ${\Pr}_{\pmb{\eta}}(X \mid \bz)$ here denotes a logistic regression model with $X$ as the outcome and $\bz$ as the covariates fit to the subset of observations with $X^* = 1$. 

\subsection{EM Algorithm}
\label{sec:intro}
To find the maximum likelihood estimator (MLE), we devise an expectation-maximization (EM) algorithm \citep{Dempster1977} based on the complete-data log-likelihood based on $(Y_i, X_i, X_i^*, \bZ_i)$ for all $N$ observations:
\begin{align}
    \ell_N(\bbeta,\boldeta) &\propto \sum_{i=1}^{N}\sum_{x=0}^{1}\textrm{I}(X_i=x)\left[\log\left\{{\Pr}_{\bbeta}(Y_i\mid X_i=x,\bz_i)\right\} + \log\left\{{\Pr}_{\boldeta}(X_i=x\mid X_i^*,\bz_i)\right\}\right]. \label{cd_loglik}
\end{align}
The indicator $\textrm{I}(X_i=x)$ is the only component in \eqref{cd_loglik} with missing data for the unqueried neighborhoods, and this missingness can be addressed through the EM algorithm. 

To start, we initialize the parameters either noninformatively $(\bbetahat^{(0)}=\pmb{0}, \hat{\pmb{\eta}}^{(0)}=\pmb{0})$ or at the complete case estimates. For the latter, $(\bbetahat^{(0)}, \hat{\pmb{\eta}}^{(0)})$ are chosen to maximize the log-likelihood among only the queried observations:
\begin{align*}
    \ell_N^c(\bbeta, \pmb{\eta}) &\propto \sum\limits_{i = 1}^N Q_i\log \{{\Pr}_{\bbeta}(Y_i \mid X_i, \bz_i){\Pr}_{\pmb{\eta}}(X_i \mid X_i^*, \bz_i)\}.
\end{align*}
The complete case estimates are expected to be consistent here, as $X$ is missing at random (MAR) under the two-phase design \citep{Little&Rubin2002}.

Parameter estimates $(\bbetahat^{(t)}, \hat{\pmb{\eta}}^{(t)})$ for subsequent iterations $t$ ($t \geq 1)$ are then updated until convergence (within $0.001$ tolerance) by the following expectation (E) and maximization (M) steps.
\begin{enumerate}
    \item \textbf{E-Step:} For unqueried neighborhoods $i$, compute the conditional expected value of the missing data in the complete-data log-likelihood, $\textrm{I}(X_i=x)$, given the fully observed data, $(Y_i, X_i^*, \bz_i)$. This expectation is defined 
    as 
    \begin{align}
        \hat{\phi}_{xi}\left(\bbeta, \boldeta\right) &= \textrm{E}\left\{\textrm{I}(X_i=x)\mid Y_i, X_i^*, \bz_i\right\} \nonumber \\
        &= {\Pr}\left(X_i=x\mid Y_i, X_i^*, \bz_i\right) \nonumber \\
        &= \frac{{\Pr}_{\bbeta}(Y_i \mid X_i=x, \bz_i){\Pr}_{\pmb{\eta}}(X_i=x \mid X^*_i, \bz_i)}{\sum_{x'=0}^{1}{\Pr}_{\bbeta}(Y_i \mid X_i=x', \bz_i){\Pr}_{\pmb{\eta}}(X_i=x' \mid X^*_i, \bz_i)}. 
        \label{E_step}
    \end{align}
    In the E-step for the $t$th iteration, the expectation in \eqref{E_step} is evaluated at the previous iteration's parameter values, i.e., $\hat{\phi}_{xi}\left(\bbeta^{(t-1)}, \boldeta^{(t-1)}\right)$. Notice that for every neighborhood $i$ ($i \in \{1, \dots, N\}$), the two expectations sum to one, i.e., $\hat{\phi}_{0i}\left(\bbeta^{(t-1)}, \boldeta^{(t-1)}\right) + \hat{\phi}_{1i}\left(\bbeta^{(t-1)}, \boldeta^{(t-1)}\right) = 1$, by the probability of the complement. 
    \item \textbf{M-Step:} Replacing the missing expectations $\textrm{I}(X_i=x)$ in the complete-data log-likelihood with their updated expectations $\hat{\phi}_{xi}\left(\bbeta^{(t-1)}, \boldeta^{(t-1)}\right)$ from the E-step admits:
    \begin{align}
        \sum_{i=1}^{N}\sum_{x=0}^{1}\hat{\phi}_{xi}\left(\bbeta^{(t-1)}, \boldeta^{(t-1)}\right)\left[\log\left\{{\Pr}_{\bbeta}(Y_i\mid X_i=x,\bz_i)\right\} + \log\left\{{\Pr}_{\boldeta}(X_i=x\mid X_i^*,\bz_i)\right\}\right], \label{cd_loglik_est}
    \end{align}
    where $\hat{\phi}_{xi}\left(\bbeta^{(t-1)}, \boldeta^{(t-1)}\right) = \textrm{I}(X_i=x)$ if neighborhood $i$ was queried. Thus, in the M-step for the $t$th iteration, the parameter values $\left(\bbeta^{(t)}, \boldeta^{(t)}\right)$ are chosen to maximize the relevant terms in the objective function \eqref{cd_loglik_est}:
    \begin{align}
        \sum_{i=1}^{N}\sum_{x=0}^{1}\hat{\phi}_{xi}\left(\bbeta^{(t-1)}, \boldeta^{(t-1)}\right)\log\left\{{\Pr}_{\bbeta}(Y_i\mid X_i=x,\bz_i)\right\} \label{weighted_poisson}
    \end{align} 
    and 
    \begin{align}
        \sum_{i=1}^{N}\sum_{x=0}^{1}\hat{\phi}_{xi}\left(\bbeta^{(t-1)}, \boldeta^{(t-1)}\right)\log\left\{{\Pr}_{\boldeta}(X_i=x\mid X_i^*,\bz_i)\right\}, \label{weighted_logistic}
    \end{align}
    respectively. Notice that \eqref{weighted_poisson} and \eqref{weighted_logistic} are the log-likelihood functions for \textit{weighted} Poisson and logistic regression models, respectively, where the expectations from the E-step are the weights. 
\end{enumerate}

\begin{figure}[ht]
\caption{The EM algorithm allows us to efficiently compute the MLE by iterating between the expectation (E) and maximization (M) steps.
\label{fig:EM}}
\begin{center}
\vspace*{-2mm}
\includegraphics[width=5in]{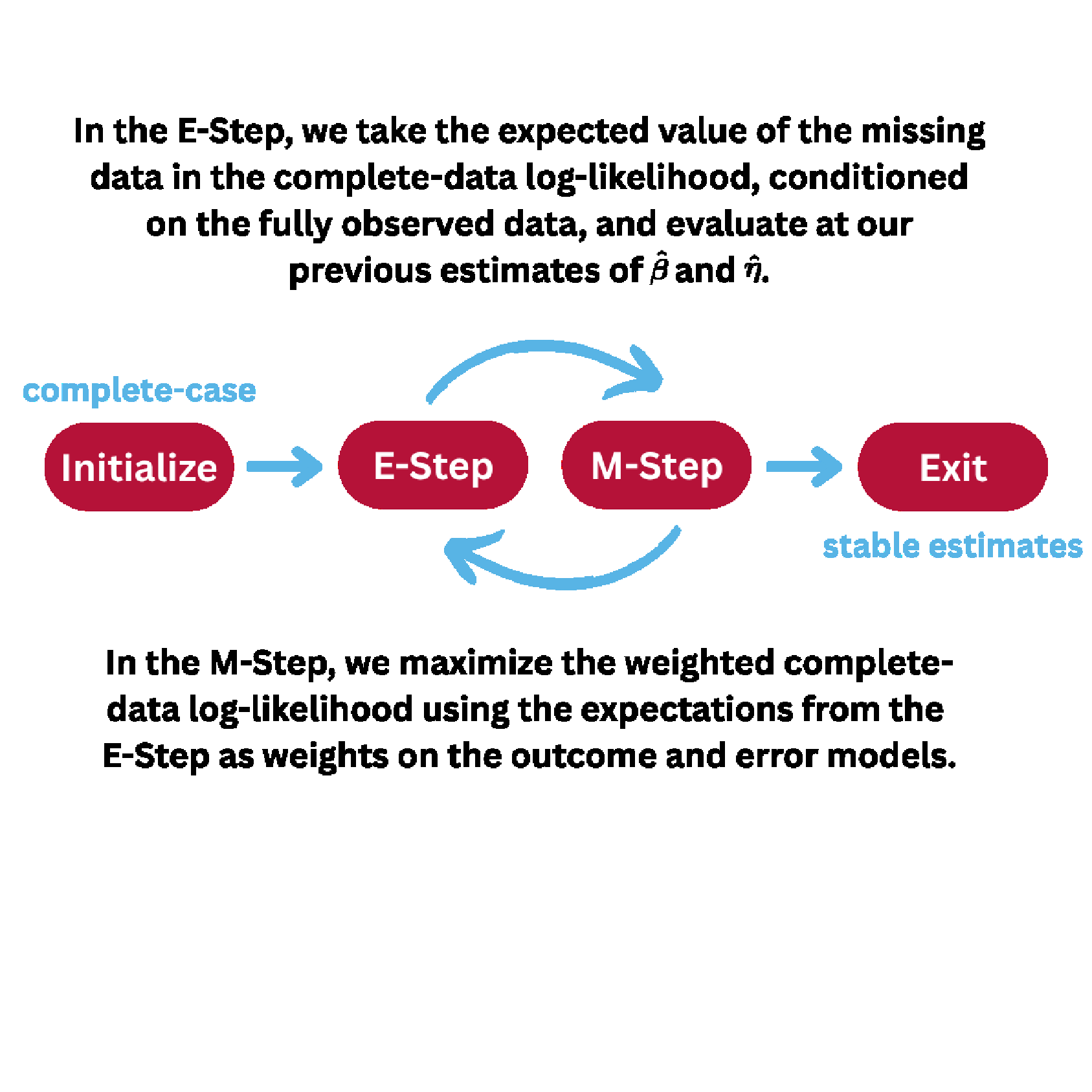}
\vspace*{-8mm}
\end{center}
\end{figure}

\clearpage

\section{Additional Details and Results from Simulation Studies}

\subsection{Data generation}\label{datagen}

First, we simulated a covariate $Z$ from a gamma distribution with shape $=0.6$ and scale $=0.2$. This covariate was chosen to represent land area (in square miles). From $Z$,the error-prone binary exposure $X^*$ was simulated from a Bernoulli distribution with probability $\Pr(X^*=1|Z) = 1/[1 + \exp\{- (1 - Z)\}]$. Next, we generated the true exposure of interest $X$ from a Bernoulli distribution given $X^*$ and $Z$ with the probability parameter: 
\begin{align}
\Pr(X=1 \mid X^*, Z) &= 
\begin{cases} 0 & \text{ if } X^* = 0 \\
\frac{1}{1 + \exp\{-(\eta_0 + 0.39 Z)\}} & \text{ if } X^* = 1 
\end{cases}.    \label{gen_X}
\end{align}
Notice that the two cases in \eqref{gen_X} induced one-sided misclassification in $X^*$, such that false positives (i.e., with $X = 0$ and $X^* = 1$) were the only possible type of errors. The intercept $\eta_0$ in \eqref{gen_X} was chosen to force a certain positive predictive value $PPV$, by letting $\eta_0 = \log\{PPV/(1 - PPV)\}$; the coefficient on $Z$ was based on the real data. In Section~3.1, we varied $PPV \in \{0.5, 0.6, 0.7, 0.8, 0.9\}$; otherwise, we assumed $PPV = 0.6$. Two-sided misclassification (i.e., with both false positives and false negatives) was considered in \ref{res:two-sided}. 

We simulated our offset $O$ (the population) from a Poisson distribution with mean $= 4165$ and the outcome $Y$ (the count of diabetes cases in a neighborhood) from a Poisson distribution with mean $\lambda = O\exp(\beta_0 + \beta_1X + \beta_2Z)$. Based on the real data, we fixed $\beta_0 = -2.28$, $\beta_1 = 0.18$, and $\beta_2 = 0.14$ for the simulations in Sections~3.1 and 3.2. In \ref{res:prev},  we varied (i) $\beta_0 \in \{-3.00, -2.30, -1.61, -1.11\}$ with $\beta_1 = 0.18$ and (ii) $\beta_1 \in \{-0.11, 0.10, 0.18, 0.41\}$ with $\beta_0 = -2.30$.

Finally, we simulated the two-phase study design. For a specified query proportion $q$, we randomly sampled $n = \nint{Nq}$ of the observations with $X^* = 1$, where $\nint{\cdot}$ denotes the nearest integer function. Any observations with $X^* = 1$ that were \textit{not} chosen to be in this subset had their $X$ values redacted (i.e., made missing) for the complete case and MLE analyses. In Section~3.2, we varied $q \in \{0.1, 0.25, 0.5, 0.75\}$; otherwise, we fixed $q = 0.1$. 

\subsection{Results under two-sided misclassification}\label{res:two-sided}

To simulate two-sided exposure misclassification, we first generated a covariate $Z$ and error-prone binary exposure $X^*$ as described in \ref{datagen}. We then simulated the true exposure of interest $X$ from a Bernoulli distribution given $X^*$ and $Z$ with the probability parameter 
\begin{align}
P(X=1 \mid X^*,Z) = \frac{1}{1 + \exp[-(\eta_0 + \eta_1X^* + 0.34Z)]}.\label{x_gen_2sided}
\end{align} The coefficient $\eta_1$ in \eqref{x_gen_2sided} was chosen to force the desired positive predictive value ($PPV$), while the coefficient $\eta_0 = -1.10$ was chosen to force a negative predictive value ($NPV$) of 0.75. We then generated the population offset $O$ and outcome $Y$ as described in \ref{datagen}. Finally, we simulated the two-phase study design by randomly sampling $n = \lfloor \frac{1}{2}Nq \rfloor$ observations with $X^*=0$ and $n = \lceil \frac{1}{2}Nq \rceil$ observations with $X^*=1$, where $\lfloor\cdot\rfloor$ and $\lceil\cdot\rceil$ are the floor and ceiling functions respectively. In this section, we fixed $q = 0.1$ and varied $PPV \in \{0.5, 0.6, 0.7, 0.8, 0.9\}$.

\begin{table}
\centering
\centering
\caption{Simulation results for $\hat{\beta}_1$, the adjusted log prevalence ratio on $X$, under varied positive predictive value ($PPV$) and sample size ($N$), assuming two-sided misclassification, $0.75$ negative predictive value, $\approx 10\%$ outcome prevalence, $\beta_1 = 0.18$, and $q =0.1$ query percentage.}
\resizebox{\columnwidth}{!}{
\begin{threeparttable}
\begin{tabular}{ccrcrcrccrcccc}
\toprule
\multicolumn{2}{c}{\textbf{ }} & \multicolumn{2}{c}{\textbf{Gold Standard}} & \multicolumn{2}{c}{\textbf{Naive}} & \multicolumn{3}{c}{\textbf{Complete Case}} & \multicolumn{5}{c}{\textbf{Maximum Likelihood Estimator}} \\
\cmidrule(l{3pt}r{3pt}){3-4} \cmidrule(l{3pt}r{3pt}){5-6} \cmidrule(l{3pt}r{3pt}){7-9} \cmidrule(l{3pt}r{3pt}){10-14}
\textbf{$\pmb{N}$} & \textbf{$\pmb{PPV}$} & \textbf{Bias} & \textbf{ESE} & \textbf{Bias} & \textbf{ESE} & \textbf{Bias} & \textbf{ESE} & \textbf{RE} & \textbf{Bias} & \textbf{ESE} & \textbf{ASE} & \textbf{CP} & \textbf{RE}\\
\midrule
390 & 0.5 & $ 0.001$ & $0.005$ & $-0.746$ & $0.011$ & $0.006$ & $0.015$ & $0.087$ & $ 0.001$ & $0.005$ & $0.005$ & $0.951$ & $0.940$\\
 & 0.6 & $-0.001$ & $0.005$ & $-0.648$ & $0.011$ & $0.005$ & $0.015$ & $0.092$ & $ 0.000$ & $0.005$ & $0.005$ & $0.948$ & $0.978$\\
 & 0.7 & $ 0.000$ & $0.005$ & $-0.551$ & $0.010$ & $0.005$ & $0.016$ & $0.095$ & $ 0.000$ & $0.006$ & $0.005$ & $0.950$ & $0.749$\\
 & 0.8 & $ 0.000$ & $0.005$ & $-0.458$ & $0.010$ & $0.002$ & $0.015$ & $0.100$ & $-0.007$ & $0.011$ & $0.005$ & $0.934$ & $0.201$\\
 & 0.9 & $ 0.001$ & $0.005$ & $-0.367$ & $0.009$ & $0.005$ & $0.016$ & $0.115$ & $-0.050$ & $0.023$ & $0.005$ & $0.812$ & $0.052$\\
\addlinespace
2200 & 0.5 & $0.000$ & $0.002$ & $-0.742$ & $0.004$ & $ 0.000$ & $0.006$ & $0.101$ & $0.000$ & $0.002$ & $0.002$ & $0.943$ & $0.973$\\
 & 0.6 & $0.000$ & $0.002$ & $-0.644$ & $0.004$ & $-0.001$ & $0.006$ & $0.092$ & $0.000$ & $0.002$ & $0.002$ & $0.950$ & $0.954$\\
 & 0.7 & $0.000$ & $0.002$ & $-0.549$ & $0.004$ & $-0.001$ & $0.006$ & $0.122$ & $0.000$ & $0.002$ & $0.002$ & $0.945$ & $0.965$\\
 & 0.8 & $0.000$ & $0.002$ & $-0.456$ & $0.004$ & $ 0.001$ & $0.006$ & $0.112$ & $0.000$ & $0.002$ & $0.002$ & $0.946$ & $0.931$\\
 & 0.9 & $0.000$ & $0.002$ & $-0.366$ & $0.004$ & $ 0.001$ & $0.007$ & $0.107$ & $0.000$ & $0.002$ & $0.002$ & $0.959$ & $0.914$\\
\bottomrule
\end{tabular}
\begin{tablenotes}[flushleft]
\item{\em Note:} \textbf{Bias} and \textbf{ESE} are, respectively, the empirical relative bias and standard error of each estimator; \textbf{ASE} is the average of the standard error estimator; \textbf{CP} is the empirical coverage probability of the \textcolor{black}{95\% Wald confidence intervals (computed with the standard error estimator)}; and \textbf{RE} is the relative efficiency of the estimator to the gold standard. All entries are based on \num{1000} replicates. 
\end{tablenotes}
\label{sims:tppv}
\end{threeparttable}}
\end{table}

\subsection{Results under varied prevalence and prevalence ratio} \label{res:prev}

Simulation results under varied outcome prevalence $\beta_0$ and prevalence ratio of the exposure $\beta_1$ are summarized in Tables~\ref{sims:beta0} and~\ref{sims:beta1}. Like the behavior described in Section 3, we see negligible bias in the MLE and dramatic gains in efficiency as compared to the complete case analysis.

\begin{table}
\centering
\centering
\caption{Simulation results for $\hat{\beta}_1$, the adjusted log prevalence ratio on $X$, under varied outcome prevalence ($\exp(\beta_0)$) and sample size ($N$), assuming $60\%$ positive predictive value, $\beta_1 = 0.18$, and $q =0.1$ query percentage.}
\resizebox{\columnwidth}{!}{
\begin{threeparttable}
\begin{tabular}{ccrcrcrccrcccc}
\toprule
\multicolumn{2}{c}{\textbf{ }} & \multicolumn{2}{c}{\textbf{Gold Standard}} & \multicolumn{2}{c}{\textbf{Naive}} & \multicolumn{3}{c}{\textbf{Complete Case}} & \multicolumn{5}{c}{\textbf{Maximum Likelihood Estimator}} \\
\cmidrule(l{3pt}r{3pt}){3-4} \cmidrule(l{3pt}r{3pt}){5-6} \cmidrule(l{3pt}r{3pt}){7-9} \cmidrule(l{3pt}r{3pt}){10-14}
\textbf{$\pmb{N}$} & \textbf{$\pmb{\beta_0}$} & \textbf{Bias} & \textbf{ESE} & \textbf{Bias} & \textbf{ESE} & \textbf{Bias} & \textbf{ESE} & \textbf{RE} & \textbf{Bias} & \textbf{ESE} & \textbf{ASE} & \textbf{CP} & \textbf{RE}\\
\midrule
390 & -3.00 & $ 0.000$ & $0.007$ & $-0.366$ & $0.009$ & $ 0.001$ & $0.022$ & $0.099$ & $ 0.000$ & $0.007$ & $0.007$ & $0.955$ & $0.936$\\
 & -2.30 & $ 0.000$ & $0.005$ & $-0.368$ & $0.007$ & $-0.004$ & $0.016$ & $0.087$ & $ 0.000$ & $0.005$ & $0.005$ & $0.951$ & $0.981$\\
 & -1.61 & $-0.001$ & $0.003$ & $-0.368$ & $0.007$ & $ 0.000$ & $0.011$ & $0.081$ & $-0.001$ & $0.003$ & $0.003$ & $0.956$ & $0.996$\\
 & -1.11 & $-0.001$ & $0.003$ & $-0.366$ & $0.006$ & $ 0.002$ & $0.009$ & $0.080$ & $-0.001$ & $0.003$ & $0.003$ & $0.957$ & $1.001$\\
\addlinespace
2200 & -3.00 & $-0.001$ & $0.003$ & $-0.368$ & $0.004$ & $-0.001$ & $0.009$ & $0.094$ & $ 0.000$ & $0.003$ & $0.003$ & $0.952$ & $0.919$\\
 & -2.30 & $ 0.000$ & $0.002$ & $-0.367$ & $0.003$ & $ 0.000$ & $0.007$ & $0.088$ & $ 0.000$ & $0.002$ & $0.002$ & $0.955$ & $0.975$\\
 & -1.61 & $-0.001$ & $0.001$ & $-0.367$ & $0.003$ & $ 0.000$ & $0.005$ & $0.083$ & $-0.001$ & $0.001$ & $0.001$ & $0.960$ & $0.998$\\
 & -1.11 & $ 0.000$ & $0.001$ & $-0.367$ & $0.002$ & $ 0.001$ & $0.004$ & $0.093$ & $ 0.000$ & $0.001$ & $0.001$ & $0.957$ & $0.999$\\
\bottomrule
\end{tabular}
\begin{tablenotes}[flushleft]
\item{\em Note:} \textbf{Bias} and \textbf{ESE} are, respectively, the empirical relative bias and standard error of each estimator; \textbf{ASE} is the average of the standard error estimator; \textbf{CP} is the empirical coverage probability of the \textcolor{black}{95\% Wald confidence intervals (computed with the standard error estimator)}; and \textbf{RE} is the relative efficiency of the estimator to the gold standard. All entries are based on \num{1000} replicates. 
\end{tablenotes}
\label{sims:beta0}
\end{threeparttable}}
\end{table}

\begin{table}
\centering
\caption{Simulation results for $\hat{\beta}_1$, the adjusted log prevalence ratio on $X$, under varied true prevalence ratios ($\exp{(\beta_1)}$) and sample size ($N$), assuming $\approx 10\%$ outcome prevalence, $60\%$ positive predictive value, and $q =0.1$ query percentage.}
\resizebox{\columnwidth}{!}{
\begin{threeparttable}
\begin{tabular}{ccrcrcrccrcccc}
\toprule
\multicolumn{2}{c}{\textbf{ }} & \multicolumn{2}{c}{\textbf{Gold Standard}} & \multicolumn{2}{c}{\textbf{Naive}} & \multicolumn{3}{c}{\textbf{Complete Case}} & \multicolumn{5}{c}{\textbf{Maximum Likelihood Estimator}} \\
\cmidrule(l{3pt}r{3pt}){3-4} \cmidrule(l{3pt}r{3pt}){5-6} \cmidrule(l{3pt}r{3pt}){7-9} \cmidrule(l{3pt}r{3pt}){10-14}
\textbf{$\pmb{N}$} & \textbf{$\pmb{\beta_1}$} & \textbf{Bias} & \textbf{ESE} & \textbf{Bias} & \textbf{ESE} & \textbf{Bias} & \textbf{ESE} & \textbf{RE} & \textbf{Bias} & \textbf{ESE} & \textbf{ASE} & \textbf{CP} & \textbf{RE}\\
\midrule
390 & -0.11 & $0.003$ & $0.005$ & $-0.398$ & $0.006$ & $-0.002$ & $0.016$ & $0.097$ & $0.004$ & $0.006$ & $0.006$ & $0.954$ & $0.802$\\
 & 0.10 & $0.001$ & $0.005$ & $-0.374$ & $0.006$ & $-0.006$ & $0.016$ & $0.091$ & $0.001$ & $0.005$ & $0.005$ & $0.941$ & $0.777$\\
 & 0.18 & $0.001$ & $0.005$ & $-0.367$ & $0.008$ & $ 0.000$ & $0.016$ & $0.088$ & $0.001$ & $0.005$ & $0.005$ & $0.954$ & $0.979$\\
 & 0.41 & $0.000$ & $0.004$ & $-0.342$ & $0.013$ & $-0.001$ & $0.015$ & $0.083$ & $0.000$ & $0.004$ & $0.004$ & $0.959$ & $1.000$\\
 \addlinespace
2200 & -0.11& $0.000$ & $0.002$ & $-0.400$ & $0.003$ & $-0.002$ & $0.007$ & $0.094$ & $0.000$ & $0.002$ & $0.002$ & $0.947$ & $0.785$\\
 & 0.10 & $0.001$ & $0.002$ & $-0.376$ & $0.003$ & $ 0.000$ & $0.007$ & $0.096$ & $0.001$ & $0.002$ & $0.002$ & $0.946$ & $0.867$\\
 & 0.18 & $0.000$ & $0.002$ & $-0.366$ & $0.003$ & $ 0.000$ & $0.006$ & $0.092$ & $0.000$ & $0.002$ & $0.002$ & $0.956$ & $0.967$\\
 & 0.41 & $0.000$ & $0.002$ & $-0.341$ & $0.005$ & $ 0.000$ & $0.006$ & $0.091$ & $0.000$ & $0.002$ & $0.002$ & $0.938$ & $1.000$\\
\bottomrule
\end{tabular}
\begin{tablenotes}[flushleft]
\item{\em Note:} \textbf{Bias} and \textbf{ESE} are, respectively, the empirical relative bias and standard error of each estimator; \textbf{ASE} is the average of the standard error estimator; \textbf{CP} is the empirical coverage probability of the 95\% Wald confidence intervals (computed with the standard error estimator); and \textbf{RE} is the relative efficiency of the estimator to the gold standard. All entries are based on \num{1000} replicates.
\end{tablenotes}
\label{sims:beta1}
\end{threeparttable}}
\end{table}

\section{Model Results at All Thresholds for Piedmont Triad Data}

\begin{figure}[ht]
\caption{In this small example of N = 25 census tracts, we conduct a two-phase design and choose a subset of n = 4 census tracts to have their food access data validated. 
\label{fig:design}}
\begin{center}
\vspace*{-2mm}
\includegraphics[width=5in]{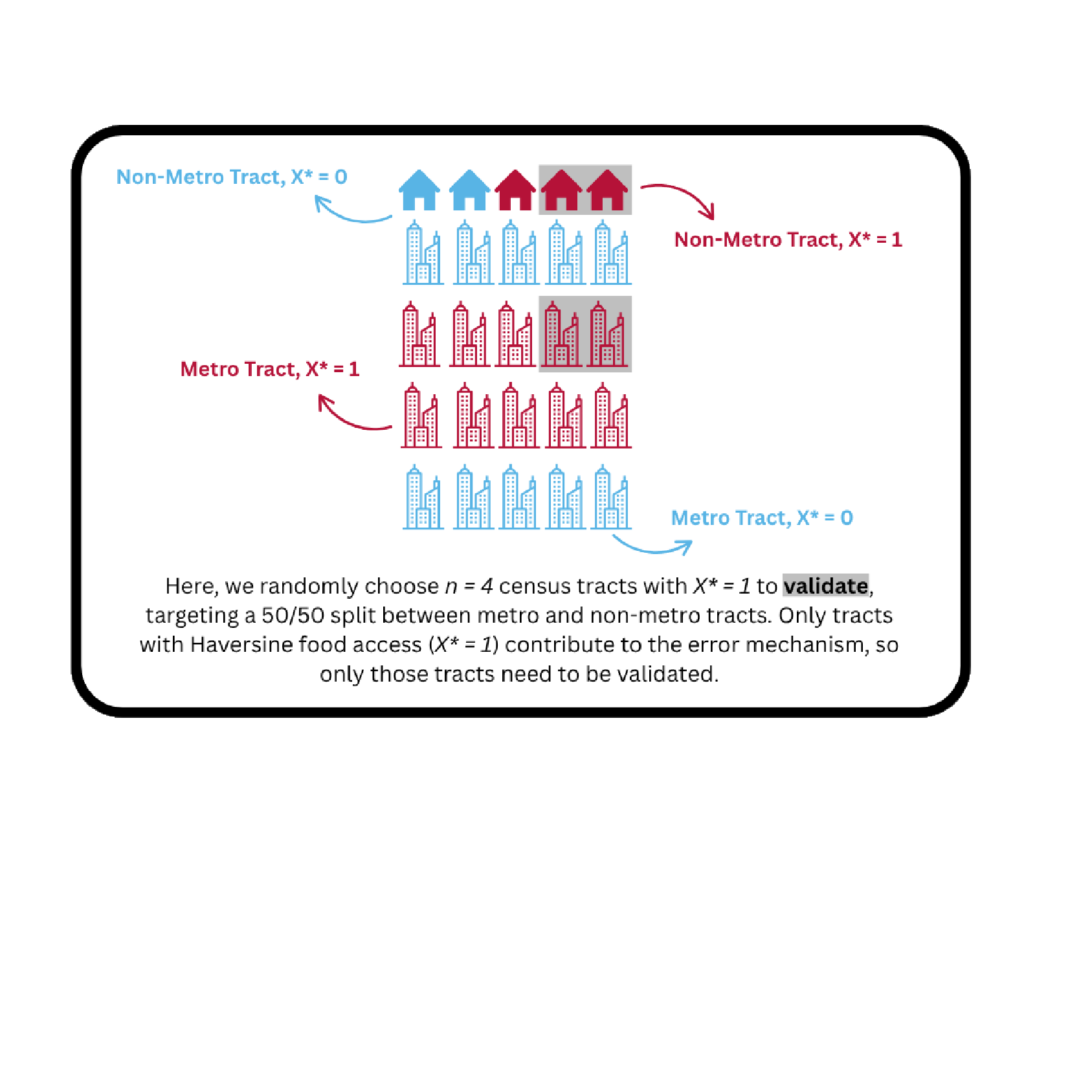}
\vspace*{-8mm}
\end{center}
\end{figure}

Figure~\ref{fig:design} provides an example of how tracts were included in the MLE model. Model results for all four analysis methods at the half-mile and one-mile thresholds are reported in Table~\ref{tab:piedmont_res}. The only notable differences between the results for the two thresholds are seen in the coefficient on the interaction term. \rev{For additional sensitivity analysis, we shrank the query percentage from the original 20\% to 10\% (Table~\ref{tab:piedmont10}). Differences were mostly negligible, although some models had slightly more extreme coefficients on the metropolitan indicator.
}

\begin{table}[ht]
\centering
\caption{Fitted models for diabetes prevalence in the Piedmont Triad based on two distance thresholds for access to healthy foods\rev{, a query percentage of 20\%,} and four analysis approaches. All models controlled for metropolitan status of the tract and included an interaction between access and metropolitan status.}
\resizebox{\columnwidth}{!}{
\begin{threeparttable}
\begin{tabular}{rrcrcrcrc}
\toprule
& \multicolumn{2}{c}{\textbf{Naive}} & \multicolumn{2}{c}{\textbf{Gold Standard}} & \multicolumn{2}{c}{\textbf{MLE}} & \multicolumn{2}{c}{\textbf{Complete Case}} \\
\cmidrule(l{3pt}r{3pt}){2-3} \cmidrule(l{3pt}r{3pt}){4-5} \cmidrule(l{3pt}r{3pt}){6-7} \cmidrule(l{3pt}r{3pt}){8-9}
\textbf{Coefficient} & \textbf{Est.} & \textbf{(95\% CI)} & \textbf{Est.} & \textbf{(95\% CI)} & \textbf{Est.} & \textbf{(95\% CI)} & \textbf{Est.} & \textbf{(95\% CI)} \\
\midrule
\multicolumn{9}{l}{\textit{Threshold: $1/2$ Mile}} \\
Intercept & $-2.12$ & ($-2.13$, $-2.10$) & $-2.11$ & ($-2.12$, $-2.09$) & $-2.11$ & ($-2.12$, $-2.09$) & $-2.11$ & ($-2.12$, $-2.09$) \\
Access & $0.10$ & ($0.08$, $0.12$) & $0.08$ & ($0.06$, $0.11$) & $0.08$ & ($0.06$, $0.11$) & $0.08$ & ($0.06$, $0.11$) \\
Metropolitan & $-0.18$ & ($-0.19$, $-0.16$) & $-0.17$ & ($-0.18$, $-0.16$) & $-0.21$ & ($-0.22$, $-0.19$) & $-0.18$ & ($-0.19$, $-0.16$) \\
Access $\times$ Metropolitan & $0.00$ & ($-0.03$, $0.03$) & $0.11$ & ($0.07$, $0.15$) & $0.14$ & ($0.10$, $0.17$) & $0.12$ & ($0.08$, $0.16$)\\
\addlinespace
\multicolumn{9}{l}{\textit{Threshold: $1$ Mile}} \\
Intercept & $-2.12$ & ($-2.13$, $-2.10$) & $-2.11$ & ($-2.12$, $-2.09$) & $-2.11$ & ($-2.12$, $-2.09$) & $-2.11$ & ($-2.12$, $-2.09$) \\
Access & $0.10$ & ($0.08$, $0.12$) & $0.08$ & ($0.06$, $0.11$) & $0.08$ & ($0.06$, $0.11$) & $0.08$ & ($0.06$, $0.11$) \\
Metropolitan & $-0.18$ & ($-0.19$, $-0.16$) & $-0.17$ & ($-0.18$,  $-0.16$) & $-0.21$ & ($-0.22$, $-0.19$) & $-0.18$ & ($-0.19$, $-0.16$) \\
Access $\times$ Metropolitan & $0.04$ & ($0.02$, $0.07$) & $0.09$ & ($0.06$, $0.11$) & $0.25$ & ($0.22$, $0.27$) & $0.09$ & ($0.07$, $0.12$) \\
\bottomrule
\end{tabular}
\begin{tablenotes}[flushleft]
\item{\em Note:} The four analysis approaches were: (i) \textbf{Gold Standard} (using queried data on all tracts), (ii) \textbf{Naive} (using unqueried data on all tracts), (iii) \textbf{Maximum Likelihood Estimator (MLE)} (using queried data on just $77$ tracts and unqueried data on the rest), and \textbf{Complete Case} (using queried data on just $77$ tracts and ignoring the others). \textbf{Est.} denotes the estimated baseline log prevalence for ``Intercept'' (i.e., expected prevalence in a non-metropolitan census tract without access to healthy foods). For all other coefficients, \textbf{Est.} denotes the estimated log prevalence ratio. \textbf{95\% CI} denotes the 95\% confidence interval. 
\end{tablenotes}
\end{threeparttable}
}
\label{tab:piedmont_res}
\end{table}

\rev{
\begin{table}[ht]
\centering
\caption{Fitted models for diabetes prevalence in the Piedmont Triad based on two distance thresholds for access to healthy foods, a query percentage of 10\%, and four analysis approaches. All models controlled for metropolitan status of the tract and included an interaction between access and metropolitan status.}
\resizebox{\columnwidth}{!}{
\begin{threeparttable}
\begin{tabular}{rrcrcrcrc}
\toprule
& \multicolumn{2}{c}{\textbf{Naive}} & \multicolumn{2}{c}{\textbf{Gold Standard}} & \multicolumn{2}{c}{\textbf{MLE}} & \multicolumn{2}{c}{\textbf{Complete Case}} \\
\cmidrule(l{3pt}r{3pt}){2-3} \cmidrule(l{3pt}r{3pt}){4-5} \cmidrule(l{3pt}r{3pt}){6-7} \cmidrule(l{3pt}r{3pt}){8-9}
\textbf{Coefficient} & \textbf{Est.} & \textbf{(95\% CI)} & \textbf{Est.} & \textbf{(95\% CI)} & \textbf{Est.} & \textbf{(95\% CI)} & \textbf{Est.} & \textbf{(95\% CI)} \\
\midrule
\multicolumn{9}{l}{\textit{Threshold: $1/2$ Mile}} \\
Intercept & $-2.10$ & ($-2.11$, $-2.08$) & $-2.09$ & ($-2.10$, $-2.08$) & $-2.09$ & ($-2.10$, $-2.08$) & $-2.09$ & ($-2.10$, $-2.08$) \\
Access & $0.11$ & ($0.08$, $0.14$) & $0.09$ & ($0.06$, $0.13$) & $0.09$ & ($0.06$, $0.13$) & $0.09$ & ($0.06$, $0.13$) \\
Metropolitan & $-0.15$ & ($-0.16$, $-0.14$) & $-0.15$ & ($-0.16$, $-0.14$) & $-0.16$ & ($-0.18$, $-0.15$) & $-0.15$ & ($-0.16$, $-0.14$) \\
Access $\times$ Metropolitan & $0.00$ & ($-0.03$, $0.03$) & $0.11$ & ($0.07$, $0.15$) & $0.29$ & ($0.25$, $0.33$) & $0.24$ & ($0.19$, $0.28$)\\
\addlinespace
\multicolumn{9}{l}{\textit{Threshold: $1$ Mile}} \\
Intercept & $-2.12$ & ($-2.13$, $-2.10$) & $-2.11$ & ($-2.12$, $-2.09$) & $-2.11$ & ($-2.12$, $-2.09$) & $-2.11$ & ($-2.12$, $-2.09$) \\
Access & $0.10$ & ($0.08$, $0.12$) & $0.08$ & ($0.06$, $0.11$) & $0.10$ & ($0.08$, $0.12$) & $0.10$ & ($0.07$, $0.12$) \\
Metropolitan & $-0.18$ & ($-0.19$, $-0.16$) & $-0.17$ & ($-0.18$,  $-0.16$) & $-0.22$ & ($-0.23$, $-0.20$) & $-0.19$ & ($-0.21$, $-0.18$) \\
Access $\times$ Metropolitan & $0.04$ & ($0.02$, $0.07$) & $0.09$ & ($0.06$, $0.11$) & $0.37$ & ($0.34$, $0.39$) & $0.12$ & ($0.08$, $0.15$) \\
\bottomrule
\end{tabular}
\begin{tablenotes}[flushleft]
\item{\em Note:} The four analysis approaches were: (i) \textbf{Gold Standard} (using queried data on all tracts), (ii) \textbf{Naive} (using unqueried data on all tracts), (iii) \textbf{Maximum Likelihood Estimator (MLE)} (using queried data on just $77$ tracts and unqueried data on the rest), and \textbf{Complete Case} (using queried data on just $77$ tracts and ignoring the others). \textbf{Est.} denotes the estimated baseline log prevalence for ``Intercept'' (i.e., expected prevalence in a non-metropolitan census tract without access to healthy foods). For all other coefficients, \textbf{Est.} denotes the estimated log prevalence ratio. \textbf{95\% CI} denotes the 95\% confidence interval. 
\end{tablenotes}
\end{threeparttable}
}
\label{tab:piedmont10}
\end{table}
}

\clearpage
\bibliographystyle{Chicago}
\bibliography{ref}